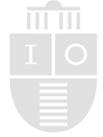

BENEDIKT FECHER, MARCEL HEBING, MELISSA LAUFER, JÖRG POHLE, FABIAN SOFSKY

# Friend or Foe?

Exploring the Implications of Large Language Models on the Science System

## ABSTRACT

The advent of ChatGPT by OpenAI has prompted extensive discourse on its potential implications for science and higher education. While the impact on education has been a primary focus, there is limited empirical research on the effects of large language models (LLMs) and LLM-based chatbots on science and scientific practice. To investigate this further, we conducted a Delphi study involving 72 experts specialising in research and AI. The study focused on applications and limitations of LLMs, their effects on the science system, ethical and legal considerations, and the required competencies for their effective use. Our findings highlight the transformative potential of LLMs in science, particularly in administrative, creative, and analytical tasks. However, risks related to bias, misinformation, and quality assurance need to be addressed through proactive regulation and science education. This research contributes to informed discussions on the impact of generative AI in science and helps identify areas for future action.

## KEYWORDS

## CITATION

## LICENCE

**AUTHOR INFO, AFFILIATION & FUNDING**

Benedikt Fecher (corresponding author), Alexander von Humboldt Institute for Internet and Society (Berlin) & Wissenschaft im Dialog (Berlin), fecher@hiig.de

Marcel Hebing, Alexander von Humboldt Institute for Internet and Society (Berlin) & Digital Business University of Applied Sciences (DBU), marcel.hebing@hiig.de

Melissa Laufer, Alexander von Humboldt Institute for Internet and Society (Berlin), melissa.laufer@hiig.de

Jörg Pohle, Alexander von Humboldt Institute for Internet and Society (Berlin), joerg.pohle@hiig.de

Fabian Sofsky, Alexander von Humboldt Institute for Internet and Society (Berlin), fabian.sofsky@hiig.de

This study was funded and primarily conducted by the Alexander von Humboldt Institute for Internet and Society. The authors and affiliated institutions declare no conflicting interests.

## CONTENTS

# 1 INTRODUCTION

The release of ChatGPT by OpenAI in November 2022 has sparked a plethora of editorials, position papers and essays, or interviews with experts, as well as some articles and preprints on the potential impacts on science and higher education. While many concerns raised relate to how ChatGPT will change education (e.g., Perkins, 2023), there is much less–especially empirical research—on the implications of large language models (LLMs) as well as LLM-based chatbots or prompts on the science system and scholarly practices (Ribeiro et al., 2023). One can however draw inspiration from fields that are also characterized by largely text-based or -focused, creative and knowledge work. For instance, the editorial by Dwivedi et al. (2023) provides a viewpoint on the potential impact of generative AI technologies such as ChatGPT in the domains of education, business, and society, based on 43 contributions by AI experts from various disciplines. However, the literature on knowledge work and the transformative effects of AI cannot account for the complexities of specific practices (Jiang et al., 2022).

In light of the limited research conducted on large language models and their impact on the science system and scientific practice, we initiated a Delphi study involving experts who specialize in the intersection of research and AI technology. The purpose of this study was to investigate the following areas: a) the potential applications and limitations in using LLMs, b) the positive and negative effects of LLMs on the science system, c) the regulatory and ethical considerations associated with the use of LLMs in science, and d) the necessary competencies and capacities for effectively utilizing LLMs. Our objective in this study was to gather and structure expert opinions in an initial phase, focusing on the aforementioned categories, and subsequently evaluate and assess them in a second phase. As generative AI continues to advance, it is crucial to gather expert knowledge and informed assessments regarding its potential impact on science. This knowledge will contribute to an informed scholarly debate and help anticipate potential fields of action.

Our findings indicate that experts anticipate that the utilization of LLMs will have a transformative and largely positive impact on science and scientific practice. In LLMs, they recognize significant potential for administrative, creative and analytical tasks. The main risks associated with LLMs pertain to issues of bias, misinformation, and overburdening of the scientific quality assurance system. Despite the perceived advantages of LLMs for science, it is imperative to acknowledge and address the associated risks. This necessitates proactive measures in regulation and science education.

# 2 LITERATURE REVIEW

In the following, we provide an overview of the current state of the scholarly discourse along the aforementioned areas. While our aim was to present a comprehensive and contemporary overview of this discourse. However, it is important to acknowledge that new and pertinent studies may have emerged by the time of the publication of this article.

## 2.1 Applications and limitations of LLMs in science

LLMs and LLM-based tools are widely expected to have a wide range of applications in scientific practice. Possible uses for researchers identified in the literature range from generating plausible research ideas (Dowling & Lucey, 2023), brainstorming (Staiman, 2023), transforming notes into text (Buruk, 2023), creating a first draft of a paper (Dwivedi et al., 2023), assisting with grammar and language (Flanagin et al., 2023), e.g. to improve clarity (Lund et al., 2023), especially for non-native speakers (Perkins, 2023), but also stylistic issues, from formatting references to complying with editing standards (Flanagin et al., 2023; Lund et al., 2023). LLM-based tools like ChatGPT may be used to generate literature reviews (Dowling & Lucey, 2023), data crunching (Staiman, 2023), data summaries (Lucey & Dowling, 2023), even proposing new experiments (Grimaldi & Ehrler, 2023). They may support the dissemination of publications and the diffusion of knowledge by helping to create better metadata, indexing, and summaries of research findings (Lund et al., 2023). They are expected to assist editors in screening submission for issues such as plagiarism or image manipulation, triaging, validating references, editing and formatting (Flanagin et al., 2023; Hosseini & Horbach, 2023). Beyond scholarly writing, LLM-based tools are expected to assist with code writing, automating simple tasks and error management (Dwivedi et al., 2023), but also in writing reports, strategy documents, emails as well as cover and rejection letters (Corless, 2023). Scientists may also use LLM-based tools for non-scholarly tasks, as a recent *Nature* poll has shown: while eighty per cent of respondents have used AI chatbots, more than half say they use them for 'creative fun' (Owens, 2023).

While the fields of application appear diverse, it is widely accepted that LLMs and LLM-based tools have limitations in scholarly use. Several editorials and Op Eds have been published that point to glaring mistakes of ChatGPT, including referencing scientific studies that do not exist (Perkins, 2023). The company behind ChatGPT, OpenAI, admits openly in its blog: "ChatGPT sometimes writes plausible-sounding but incorrect or nonsensical answers" (OpenAI, 2022). At the time of writing this article, all existing LLM-based chatbots have been trained on outdated data. As a result, they do not possess the capability to incorporate real-time data automatically, leading to a lack of updated information (Dwivedi et al., 2023). Other limitations that have been identified include flawed logical argumentation, lack of critical elaboration, and unoriginal generated content (Dwivedi et al., 2023). Errors may also occur in interpreting meaning, in particular if terms are ambiguous, have multiple meanings or consist of compound words (Lund et al., 2023). In addition, generated texts may lack semantic coherence and lexical diversity (Perkins, 2023). Teubner et al. (2023, p. 96) state that the produced texts often "read somewhat bland, generic, and vague with a noticeable tendency to seek balance", and that a very common ChatGPT phrase is: "However, it is important to note…". Like ML-based systems in general, LLM-based chatbots are considered to lack transparency and explainability (Dwivedi et al., 2023), and reproduce or even amplify biases inherent in the information that was used to train them (Corless, 2023; Hosseini et al., 2023), reproducing an "of the same old trivialities and stereotypes" (Teubner et al., 2023, p. 99). This is considered a structural issue of how these systems are trained and cannot be resolved by simply creating bigger models as size does not guarantee diversity (Bender et al., 2021).

## 2.2 Opportunities and risks for the science system

A prevailing viewpoint in the literature anticipates positive effects of LLMs on the science system. Potentially opportunities of LLMs on science include positive effects on scholarly productivity, quicker access to available scholarly resources via enhanced search engines to the automation of mundane, repetitive or tedious work such as correcting grammatical errors, allowing people to focus on creative and non-repetitive activities (Dwivedi et al., 2023; Lund et al., 2023). Foremost among these anticipated benefits is the enhancement of research productivity and the elevation of publication quality. There is an expectation that by using these tools to improve their texts, researchers "can focus more on what to communicate to others, rather than on how to write it" (Pividori & Greene, 2023, p. 15). Staiman (2023 n.p.), for instance, notes that the writing process should be considered less an end in itself but rather "a means to an end of conveying important findings in a manner that is clear and coherent". Along these lines, Lund et al. (2023) suggest that the capability of ChatGPT and the like might lead to questioning the strong belief that 'publish or perish' is an important and valuable principle in academia and possibly change the criteria for evaluating tenure. Some scholars expect a revolution of "the whole scientific endeavor" and refer to these tools' fundamental disregard of the boundaries of scientific disciplines, which may help "bringing multidisciplinary science to new heights" (Grimaldi & Ehrler, 2023, p. 879). Furthermore, these tools may also lead to the democratization of science: First, the research process might be democratized as LLM-based tools may compensate for the lack of financial resources, e.g. for "traditional (human) research assistance" (Lucey & Dowling, 2023 n.p.). Second, the dissemination of knowledge might be democratized as these tools can easily polish the language of a text or even translate research output to multiple languages, both of which would level the field for researchers who speak English as a foreign language (Corless, 2023; Liebrenz et al., 2023).

Among the risks for the science system identified in the literature are the adverse effects on the academic quality assurance mechanisms and, subsequently, on scientific integrity. The avalanche of AI-generated "scientific-looking papers devoid of scientific content" (Grimaldi & Ehrler, 2023, p. 879) is widely expected to overburden the academic review process and foster plagiarism (Dwivedi et al., 2023). Biases are expected to be reinforced and errors introduced into the scholarly debate that might be difficult to identify and correct (Lund et al., 2023). A recent study by Liang et al. (2023) evaluating the performance of several widely-used GPT detectors found that they consistently misclassify non-native English writing samples as AI-generated, whereas native writing samples are accurately identified. Several scholars expect that LLMs may lead to an increase in misinformation and disinformation and more "junk science" (Corless, 2023 n.p.). In this regard, Lund et al. (2023) worry that the use of LLM-based tools in academia not only raises concerns about the reproducibility and transparency of research but may undermine trust in the scientific process (see also Van Noorden, 2022).

### 2.3 Competencies and capacities in scientific practice

It is assumed that LLMs and LLM-based tools will mark a shift in the academic skill set. Prompt engineering, developing and producing prompts for conversational AI systems like ChatGPT or is the most discussed new competence that is required from researchers (Teubner et al., 2023). This is believed to pose a particular challenge for individuals who already struggle with basic IT, as they will not derive much benefit from advances in AI, and this may lead to a widening productivity gap. As LLM-based tools may have better English writing skills than some people, especially non-native speakers, the focus in academic work is expected to shift from text writing to conducting research, which requires researchers to formulate interesting research questions and carry out research to find answers (Dwivedi et al., 2023). More generally, as Teubner et al. (2023, p. 98) observe, "the ability to read and interpret different text options becomes more important than the ability to write them." That means that researchers must be able to check the generated text for factual and citation accuracy, bias, mathematical, logical, and commonsense reasoning, relevance, and originality (Hosseini et al., 2023). That also means that researchers are expected to have the competencies to collate and combine the results that LLM-based tools generate (Floridi & Chiriatti, 2020). Not surprisingly, Dowling & Lucey (2023) find that adding domain expertise greatly improves the quality of the generated results. Thus, among the key skills that researchers have to develop are critical thinking, problem solving, ethical decision-making, and creativity (Dwivedi et al., 2023).

### 2.4 Ethical and regulatory issues concerning ChatGPT

The existing literature often frames negative implications, i.e. risks, for the science system as ethical issues, and also mixes ethical and legal aspects. Issues are raised on how we understand 'authorship' in the research context, be it as accountability, as a substantial contribution to a text, as ownership in contrast to plagiarism, and with respect to text and language improvement (Staiman, 2023). Critics argue that chatbots cannot take responsibility for the content they produce and cannot be held accountable (Corless, 2023; Liebrenz et al., 2023). In addition, their ability to generate quality academic research ideas "raises fundamental questions around the meaning of creativity and ownership of creative ideas" (Lucey & Dowling, 2023 n.p.), which in turn sparks questions about originality, scholarly citation practices and the boundary to plagiarism (Lund et al., 2023; Tomlinson et al., 2023). It thus comes as no surprise that publishers like *Springer Nature* have banned ChatGPT and similar software from being given authorship on papers: and *Science* editors have also prohibited the use of any text generated by those tools. Many commentators have raised concerns about the implications of the LLMs producing inaccurate or misleading output and the potential spread of misinformation (Dwivedi et al., 2023; Liebrenz et al., 2023). Similar ethical concerns are raised regarding the potential of these tools to reproduce and amplify bias, both in the training data and the development process, and the implications of this for the integrity of science (Lund et al., 2023). There have been techno-solutionist claims that potential harms of such systems can be mitigated by watermarking their output (Kirchenbauer et al., 2023). Additional ethical considerations include the potential to replace humans in the scholarly work process (Lund et al., 2023). This includes positions that were thought to be less likely to be automated until a few years ago (Dwivedi et al., 2023). Furthermore, the commercialization of these tools would exclude scholars and institutions in low-income and middle-income countries, thus entrenching existing inequalities in knowledge dissemination and scholarly publishing (Liebrenz et al., 2023).

There is a broadly perceived lack of regulation, or at least clear regulatory guidance for LLMs and related tools, on issues such as privacy, security, accountability, copyright violations, disinformation, misinformation and other forms of abuses and misuses of LLMs and LLM-based tools (Dwivedi et al., 2023; Khowaja et al., 2023; Lund et al., 2023). After the Italian Data Protection Authority imposed an immediate temporary limitation on the processing of Italian users' data by OpenAI in late March 2023 in order to enforce demands on the protection of data subjects' rights (GPDP, 2023), other national data protection authorities in Europe have followed suit and opened proceedings against OpenAI (Sokolov, 2023). European data protection authorities have even set up a task force to cooperate and exchange information on enforcing EU laws on OpenAI (Goujard, 2023). At the same time, the European Parliament called for expanding the potential reach of the proposed EU AI Act by including ChatGPT-like systems to the list of high-risk categories of AI systems (Helberger & Diakopoulos, 2023). Furthermore, Hacker, Engel & Mauer (2023) call for specific regulation of LLM-based tools, "large generative AI models", under the EU Digital Services Act and provide four concrete, workable suggestions that include transparency obligations, mandatory yet limited risk management, non-discrimination data audits, and expanded content moderation.

## 3 METHODOLOGY

To address our research objective, we employed the Delphi method. First developed in the 1960s, the Delphi method is a technique used to establish consensus among a group of experts on complex issues (Landeta, 2006) and in some cases used to forecast future developments (Linstone & Turoff, 1975). In its basic form, this method can be described as a communication process that involves engaging experts at various stages, such as through surveys and qualitative interviews. The initial stage is open and exploratory, with the information gathered analyzed and used to inform subsequent data collections. This process continues until consensus is reached among experts, for example, in defining concepts and/or trends or weighing different viewpoints. In this light, the Delphi method is a fitting technique to investigate our objective of exploring the impact of ChatGPT and LLMs on scientific practices and the science system.

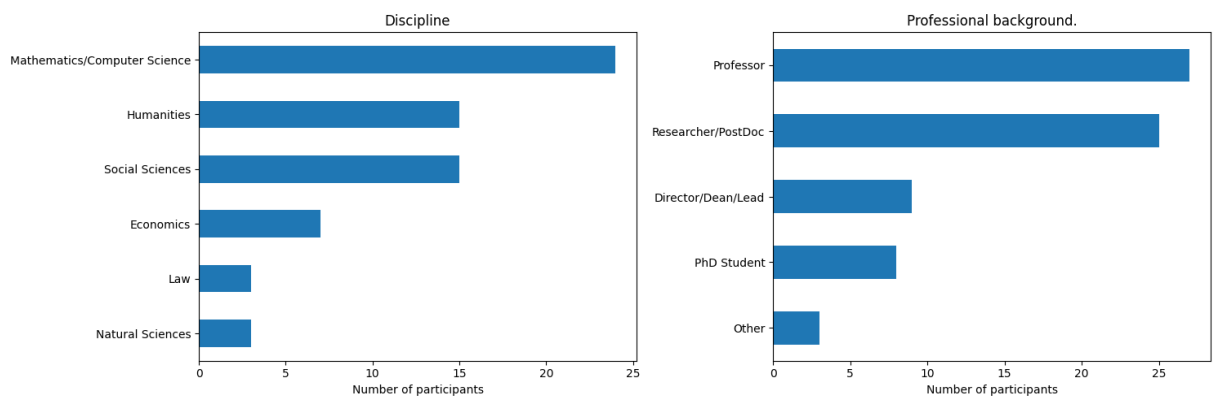

*Figure 1. Initial sample (round one) of our delphi study. Overview of participants by discipline and professional status (n=72).*

The responses of the first survey were primarily coded by two authors. They first examined 25% of the responses to generate a codebook through a combination of inductive and deductive coding (Bazeley, 2009). The codebook was then evaluated by all authors and adjustments made when needed and the rest of the material was coded (for codebook see appendix table 7 and 8). Based on this analysis, the second survey was created consisting of 11 questions, the majority of which were ranking questions featuring the identified codes for applications and limitations, risks and opportunities for the science system and the competencies needed for using LLMs, as well as general opinion questions on LLMs impact on science and scientific practice. Furthermore, the survey instrument contained two open questions on future scenarios that we analyzed for the discussion part of the paper.

The survey was sent to the same experts, yielding 52 responses (72 % of the participants from the first round). A statistical analysis was conducted on the opinion and ranking questions. In the result tables (see tables 2 to 6 in the appendix), we provide the individual frequencies for each item and rank, as well as two scores. The first score (sum) is a simple sum of the preceding frequencies, the rank is a weighted sum, where the first rank is weighted by factor four and the second rank by factor two. The rank questions are followed by a set of statements, which the participants could evaluate on a five-point likert scale (strongly disagree, disagree, neither agree nor disagree, agree, strongly agree). We combined agree and strongly agree to sort the items and will also refer to the combination of both, when reporting it in the text. The open questions were analyzed with a combination of inductive and deductive coding, carried out jointly by the authors. Our delphi approach allowed us to identify and refine various implications of LLMs on the science system, however it was not without its limitations. For example, we were unable to track long-term implications as the interval between the data collections were relatively short.

We sought consent prior to each survey phase to publish the responses, aiming to enhance the transparency of our results and enable future research and educational use. The data (including the survey instruments) is published under a CC-BY-license and can be accessed via the following [link](link).

## 4 RESULTS

Below, we present the Delphi study results based on the defined aspects, i.e. applications and limitations, risks and opportunities for the science system, competencies as well as legal and ethical implications. In each section, we begin by presenting the coded findings from phase one and use the results of the ranking and opinion questions to contextualize and weigh these results, when applicable. Figure 2 displays the results of the opinion questions, which we will refer to in the subsequent result sections. The results of the ranking questions analysis can be found in the appendix (tables 2 to 6).

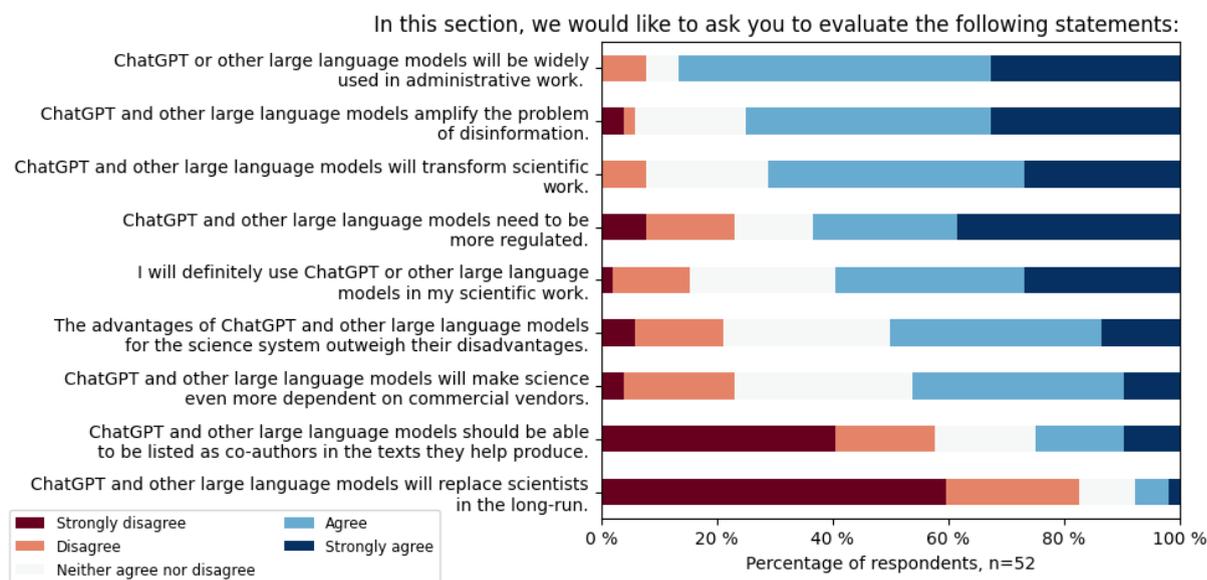

*Figure 2. Statements on LLMs, formulated based on the results from the first round of our delphi study and quantified in the second round.*

## 4.1 Applications and limitations in use: LLMs as enhancement tools

The first phase yielded six distinct applications that can be effectively addressed by LLMs and LLM-based applications. These include (1) *text improvement*, which involves the rephrasing and optimization of textual content, (2) *text summary*, which involves the summarization of information, (3) *text analysis*, such as the use of sentiment analysis or qualitative coding, (4) *code writing*, which involves assistance in programming tasks, (5) *idea generation*, which involves generating new ideas through the combination of concepts, and (6) *text translation*, which includes the translation of a text entered into the LLM in different languages. Notably, the identified applications of LLMs extend beyond conventional text-based tasks in scientific publishing, although such tasks remain a dominant practice in the responses.

In the second round of the Delphi survey, we asked the experts to prioritize the identified applications. Our results show that *text improvement* is considered the most important application, followed by *text summary* as the second most important, and *code writing* as the third most important application. Most (59.6%) of the experts either already use or express their intention to use LLMs in their own work (figure 2). A significant portion (86.5%) of the experts perceive LLMs as valuable for administrative tasks, confirming the assumption that time savings are expected for researchers through LLM utilization (figure 2).

Asked about the limitations of LLMs in scientific work, five distinct types of limitations were mentioned. We observed (1) *lack of transparency*, as it is unclear on which data the model's outputs are based on, (2) *incorrectness*, especially regarding literature references and biographical information, which may affect the reliability of the generated text, (3) *lack of creativity*, as ChatGPT relies heavily on existing patterns and may struggle to generate entirely new content, (4) *outdatedness*, particularly as the version of ChatGPT used in this study relies on a database that only goes up to 2021, and (5) *unspecificity*, i.e. LLMs produce superficial texts that do not address topics in depth or detail. There are already approaches to address some of these limitations, even if not completely.

In the second round of the study, participants were requested to rank the limitations. The highest-ranked limitation was *incorrectness*, followed by *non-transparency* and *unspecificity* in the responses. The incorrectness of LLMs was a dominant and recurring issue mentioned by the experts. As one expert stated, "The largest problem I see are the factual mistakes, often given with confidence, which make it hard to trust ChatGPT and similar technology outputs without further research or prior knowledge".

The results indicate that the potential benefits of LLMs lie not only but primarily in text-based work, which is significant because scientific value creation in most disciplines is text-based. There is also evidence to suggest that LLMs are relevant for ideation, conception, and programming, the latter of which is an increasingly important scientific practice. Taken together, it is not surprising that a majority of the respondents assume that ChatGPT and other LLMs will transform scientific practice, although this might – at this stage – relate primarily to the textuality of academic work. The limitations mentioned can be essentially explained by the databases that existing LLMs were trained on, and it can be assumed that many of these limitations can be addressed in newer models, as some respondents pointed out. However, the non-transparency in the training data remains problematic and was viewed by some as inconsistent with scientific principles of quality.

### 4.2 Risks and opportunities for the science system: advantages trump disadvantages

According to the experts, the use of LLMs provides the science system with four opportunities: (1) LLMs can promote *efficiency* by automating and supporting text work, (2) LLMs may promote *reflection* by identifying biases and new research areas, (3) LLMs may *reduce administrative workload*, (4) LLMs can promote *inclusiveness* by leveling the playing field between researchers from different backgrounds and institutions, such as those who lack resources for grant writing or those who are non-native English speakers, and (5) LLMs promote *productivity* by freeing up time for researchers to conduct more analyses or produce more scientific articles. In the second phase of the Delphi study, the experts ranked these, with the *reduction of administrative tasks* ranked first, followed by more *efficiency* and *inclusiveness* (see table 4). These results indicate that researchers see LLMs primarily as a tool to relieve and simplify their workload. Hence, a large majority of the experts disagrees that LLMs could replace researchers (82.7%, figure 2).

The analysis of the first phase of the Delphi study reveals the existence of seven distinct risks associated with the use of LLMs in scientific work. These risks include (1) *reinforce bias / dominant voices*, because statistical systems favor mainstream opinions, (2) *overburden academic quality assurance mechanisms* with semi-automated papers, (3) reinforce *inequalities* between researchers who have access to LLMs and those who do not, (4) increase *dependence* on commercial providers, (5) encourage academic *misconduct*, either intentional or unintentional by researchers, (6) lead to a *decrease in originality* due to the generic nature of LLM-generated text, and (7) the possibility to an increase in *disinformation*, which could potentially challenge scientific truths in the public domain. In the second phase, the experts ranked these, indicating that *bias* is seen as the biggest threat, followed by *disinformation* and *overburdening academic quality assurance mechanisms* (see table 5).

These risks are significant as they touch on fundamental pillars of scientific ethics and good practice, such as scientific freedom regarding the dependence on commercial publishers, scientific quality assurance concerning the handling of highly generic publications, as well as the public legitimation of science, which could be put into question by plausible and seemingly scientific nonsense produced by LLMs – large majority of the experts (75.0%) regard LLMs as a catalyst for disinformation (figure 2). Notwithstanding the gravity of the aforementioned risks, the majority of experts perceive the benefits of LLMs to outweigh the drawbacks (figure 2), which explains why most of them already use or intend to use LLMs in their work. This, however, can also be attributed to the sampling strategy employed in this study, possibly involving technology-proficient experts. This result is noteworthy nonetheless and supports the hypothesis that generative AI will change scientific work in the long run.

### 4.3 Competencies in usage: scientists need to learn to (re)think

In the inquiry regarding the competencies required for researchers to utilize ChatGPT and other LLMs, the respondents pointed out four distinct competencies, namely (1) *technical know-how* to comprehend the inner workings of LLMs, (2) the *ability to contextualize results* utilizing the outcomes generated by LLMs in practical scenarios, (3) a *reflective mindset* to consider the feedback effects on scientific practice, and (4) *ethical understanding* to responsibly employ LLMs. In the second phase, they ranked a *reflective mindset* first, followed by the *ability to contextualize results* and *ethical understanding*. The results indicate that the experts anticipate feedback effects on science, while also suggesting that the responsible application of knowledge will become even more paramount in the future.

It can be argued that reflexivity highlights the ethical implications of AI on scientific practices and ways to proactively address them, while contextuality focuses on the practical use of AI-supported findings and strategies for maximizing their utility. Our findings suggest that generative AI should be incorporated in scientific training and science education, specifically in relation to scientific ethics and effective communication of AI-driven results in their appropriate context.

### 4.4 Ethical and legal implications: clear need for regulation

The answers in the first phase allow to discern five ethical implications, namely (1) the *need for accountability in relation to the outcomes produced by LLMs*, (2) the question of *originality with regards to human creativity* (e.g., concerns of plagiarism arise), (3) the *sustainability issue* regarding the environmental effects of LLMs, (4) the *potential exclusion of researchers* who lack access to LLMs, raising concerns about universalism, and (5) the *issue of autonomy*, in which researchers may become overly dependent on (commercial) AI tools. The comments show clearly that the majority deem ChatGPT unfit for authorship due to its inability to assume responsibility for the results.

The experts perceive legal implications regarding (1) *copyright*, due to the unclear infringement of intellectual property by LLMs, (2) *data protection*, due to the ambiguity of the data used and how OpenAI utilizes input data, and (3) *liability*, due to the uncertainty of the extent to which LLMs can be held responsible for criminal errors. A large majority of the experts (63.5%) believe that LLMs should be subject to stronger regulations (figure 2).[1]

The initial round of the Delphi survey revealed that the ethical implications discussed frequently underscore the significance of the human element in scientific endeavors. This includes the responsibility and accountability of individuals for their contributions, the value of creativity and generating novel ideas, ensuring equitable access to science and the scientific community, and addressing the potential risk of dependency on LLM-based tools that may hinder individual skills and capabilities in scientific work. The amount of energy that is necessary both for training models and running inference and the CO2 footprint are mentioned as primary examples for the ecological sustainability issues ChatGPT and the like present. Taking into account that LLMs are trained on works produced by others and produce (or co-produce) works, both of which almost certainly fall under copyright law, it is not surprising that a large majority of the experts identify issues with copyright law as a pressing legal implication. The lack of transparency regarding the personal data on which the LLMs were trained, but also the further possible uses of personal data generated by the use of the tools, certainly explains why many respondents identify privacy and data protection law issues as considerable legal challenges. Whereas accountability is identified by many experts as a key ethical challenge, this does not carry over to the legal principle of liability that builds on it, which is mentioned by relatively few respondents.

---

[1] We did not ask the participants to rank the ethical and legal implications in the second round of the Delphi survey.

### 4.5 Transformative and deformative scenario

In the first phase of the Delphi, we consulted with experts to ascertain the potential impact of LLMs on scientific practice within the next 5-10 years. In the subsequent phase, we investigated the potential influence of generative AI on the relationship between science and society. Based on the answers to these questions, our study reveals two possible scenarios, namely (1) a utopian transformative scenario and a (2) dystopian deformative scenario. It is noteworthy that the negative scenario is almost a negation of the positive scenario and vice versa. However, overall, there are significantly more indications (in terms of the number of codes) for a positive scenario, which was also confirmed by the opinion battery in Phase 2.

In the utopian scenario, integrating generative AI into scientific practices offers transformative potential, overcoming path dependencies in scientific practice and accelerating scientific and societal progress. Our analysis identifies three key aspects of its impact on science: (1) streamlining repetitive tasks, (2) promoting inclusivity, and (3) facilitating interdisciplinary research. The experts propose that generative AI could automate administrative and generic tasks, freeing up time for critical reflection, analysis and innovation. It may democratize access to scientific resources, foster diversity of voices and collaboration, and aid in discovering connections across different schools of thought. The integration of generative AI tools aligns research with societal challenges, driving technological development and supporting evidence-based decision-making. Effective science communication and education are enabled through AI-driven tools. This collaborative approach propels scientific advancements towards innovative solutions.

In a dystopian scenario, the anticipated positive impacts of generative AI are largely negated, as our analysis reveals three crucial aspects: (1) a decline in research quality due to plausible yet flawed results, compromising reliability and validity; (2) a loss of research diversity through amplifying mainstream voices, resulting in missed opportunities for novel perspectives; and (3) a decrease in scientific integrity, as the ease of producing AI-generated content raises risks of reinforcing predatory publishing practices and disseminating false information, leading to confusion and distrust. The perpetuation of plausible nonsense could further have negative consequences for society when policy decisions or public opinions rely on unreliable information. Additionally, dependence on commercial providers for generative AI tools raises concerns among experts about the lack of independence and control over scientific research, potentially leading to conflicts of interest and biases in research results. Furthermore, the loss of diversity in research and a decrease in scientific integrity perpetuate biases, eroding credibility and leading to conflicts of interest, ultimately distorting the pursuit of knowledge.

## 5 DISCUSSION & CONCLUSION

The aim of this study was to investigate the impact of ChatGPT and other LLMs on the science system and scientific practices by examining their potential applications, limitations, effects, ethical and legal considerations and the necessary competencies needed by users. To date, scholars have primarily focused on the implications of LLMs on education (e.g. Perkins, 2023) with limited attention being paid to their impact on science and scientific practices (for exception, see Ribeiro et al., (2023). The overnight popularity ChatGPT experienced since its debut in November 2022 stressed even more the necessity to evaluate the implications of LLMs for science and scientific practice. To examine these implications, we employed a two-stage Delphi method, which included inviting experts, researchers working in the fields of science, technology and society to participate in two surveys as means to identify and refine the impact of LLMs on the science system and scientific practices.

At the time of the second round of our Delphi method, less than half a year had passed since the first preview of ChatGPT. Accordingly, it is difficult to make concrete predictions about the potential capabilities of future versions of LLMs like ChatGPT. Nevertheless, our study presents a consistent picture from experts which furthers our understanding of future expectations of LLMs. We were also able to identify patterns emerging regarding potential opportunities and risks. It is important to note the majority of the experts saw no danger that LLMs will replace the traditional scientist in the foreseeable future.

Overall, the experts in our study were optimistic and in agreement that the advantages of this technology outweigh their disadvantages. This optimism was paired with thoughtful concerns, which allow us to paint a nuanced picture of the potential positive and negative implications of LLMs. In general, ChatGPT and other LLMs were collectively understood as potential 'time-savers' to be used to improve and streamline the writing process, especially academic writing. For example, text improvement as in the rephrasing and optimization of textual content was considered the most important application. This outcome resonates with the scholarly discourse which highlights how generative AI can be used to enhance texts, such as with brainstorming (Staiman, 2023), crafting literature reviews (Lucey & Dowling, 2023), and improving text clarity (Lund et al., 2023). At the same time, experts in our study were aware of the limitations of LLMs and cited similar apprehensions to those raised in the literature (Dwivedi et al., 2023; OpenAI, 2022; Perkins, 2023). The experts highlighted key shortcomings such as AI produced texts may have incorrect information, their origin and referencing is non-transparent and that they lack specificity, shortcomings which are at odds with the principles of good scientific practice. It is not surprising that our study reinforced text-based applications and limitations for LLMs identified in the scholarly discourse, as text production is a key scientific practice. However, this focus may shift in the future as more usages of LLMs are explored.

In addition, our study indicates that LLMs have the potential to reshape the science system. The experts anticipate that they will lead to more efficient workflows, with the reduction of administrative tasks being ranked the highest anticipated change. This forecast supports claims made by other scholars, who argue that LLMs will help automate mundane tasks and free up space for creative thinking (Lund et al., 2023; Dwivedi et al., 2023). Other changes LLMs bring to the science system are however more complex. For example, our findings point to a double-edged sword embedded within the LLM constellation: this technology could serve to both promote inclusion and reinforce biases. On the one hand, LLMs can level the playing field for non-English speakers as they can provide editorial support, but on the other hand, they can also increase inequalities by drawing on mainstream opinions and widening the gap between those who have access to these technologies and those who do not. This multifaceted concern was also echoed by other scholars (Corless, 2023; Liebrenz et al., 2023).

The most pressing fear we identified is that LLMs perpetuate disinformation and will overburden quality assurance mechanisms in academia. In other words, LLMs will increase the sheer quantity of potentially incorrect papers and the peer-review process will simply be unable to keep up with the volume leading to a drop in quality. Similar thoughts are discussed by other scholars (Grimaldi & Ehrler, 2023; Lund et al., 2023), with these changes being described in revolutionary terms in which LLMs are positioned as the great 'game-changers' of academia. In contrast, experts in our study were more cautious with such claims seeing these changes as more incremental and pragmatic.

Moreover, our study provided insights into the competencies researchers need to be able to utilize LLMs. In line with scholars such as Teubner et al. (2023), experts in our study voiced concerns that ChatGPT and other LLMs have the potential to widen the digital divide between researchers who possess technical know-how and researchers who do not. Furthermore, the experts pointed out that the researcher's role in the writing process will shift from being the originator of ideas and texts to being required to contextualize and reflect on AI generated results. This change will entail a new way of thinking about key scientific practices and the role the individual academic plays in them. Our experts also expressed the importance of researchers having an ethical understanding, e.g. using AI in a responsible manner. A point that was only marginally addressed in the literature (Dwivedi et al., 2023). Underlying these findings is the understanding that it is up to the individual academic to ensure that they have the skills and knowledge needed to navigate these technological changes. Such a stance can contribute to furthering digital divides due to preexisting uneven digital literacy between academics, institutions and higher education systems.

Our study aided in disentangling the ethical and legal implications of ChatGPT and other LLMs. The findings further articulate the issue of authorship when it comes to using AI, an issue discussed by other scholars (Lucey & Dowling, 2023; Tomlinson et al., 2023). The majority of the experts deem that ChatGPT cannot claim authorship due to its inability to assume responsibility for its actions. In this light, the experts centred on distilling the role humans play in being accountable for their usage of LLMs, taking into consideration issues such as plagiarism, copyright and data protection. Thus, they underlined that human responsibility in AI usage is both a legal and ethical challenge, a sentiment that echoes the arguments of critics who postulate that chatbots cannot take responsibility for their actions (Corless, 2023; Liebrenz et al., 2023). In addition, the issue of access was highlighted as an ethical dilemma, that is, not all researchers will have equal access to such technologies, potentially furthering inequalities. Furthermore, the $CO_2$ emissions generated by these use of AI technologies poses environmental risks (Hao, 2019). The complexities of these ethical and legal implications show the need to take diverse issues into account when it comes to regulating the usage of LLMs in academia.

Lastly, our study presents potential future pathways for AI and its impact on the science system and society in the form of future scenarios constructed from our data. In the positive transformative scenario, the integration of LLMs in scientific practice holds great potential for improving scientific productivity, efficiency, education, communication, creativity, and discovery. In other words, LLMs can automate repetitive tasks, allowing researchers to allocate more time and resources to analytical and innovative work. It is the prevailing perception of the experts that suggests that this scenario is more likely to occur. However, it is crucial to acknowledge the potential negative deformative scenario. Experts raised concerns about the impact of generative AI on scientific quality, integrity, and the scientific ecosystem. Issues such as decreased scientific rigor, reproducibility, and a potential homogenization of science were highlighted. In addition, the reliance on generative AI models without proper validation may lead to a decrease in critical thinking and creativity.

We can strive to ensure the positive scenario by addressing the concerns highlighted in our study. In this regard, striking a balance between embracing the benefits of LLMs and upholding scientific principles is crucial. Accordingly, we should remember our scientific tools, the good practices of scientific work, and create appropriate frameworks and conditions that enable us to make use of the diverse opportunities these technologies might have to offer. At the same time, we must withstand any attempt to compromise the quality standards that we as a science community have established and which distinguishes the scientific discourse. Researchers and policymakers should invest in transparency, accountability, and comprehensive validation processes to maintain the credibility of scientific research. Efforts should be made to address biases, ensure diversity, and guard against the potential misuse of generative AI. Additionally, ongoing training and education on the responsible use of LLMs are essential for scientists to adapt to the evolving scientific landscape.

In conclusion, while the transformative scenario holds great promise for the positive impact of LLMs on the science system and society, it is imperative to proactively address the potential risks and challenges to ensure that the integration of generative AI in science is guided by ethical considerations, scientific integrity, and a commitment to societal benefit.

## 6 ACKNOWLEDGEMENTS

We express our gratitude to the participating experts, the majority of whom have given their consent to be named in the second phase of the Delphi. These experts are Alexander Terenin, Alison Kennedy, Anaëlle Gonzalez, André Vellino, Andrea Klein, Benjamin Tan, Brigitte Mathiak, Christian Gagné, Christian Vater, Daniel Guagnin, Debora Weber-Wulff, Ekaterina Hertog, Eva Seidlmayer, Evgeny Bobrov, Fabro Steibel, Fiona Kinniburgh, Florian Hoffmann, Georg von Richthofen, Graham Taylor, Hadi Asghari, Hendrik Send, Ingrid Richardson, Johannes Breuer, Katharina Mosene, Klaus Gasteier, Marina Gavrilova, Mark Spektor, Martin Schmidt, Maximlian Heimstädt, Mike Thelwall, Naireet Gosh, Natalie Sontopski, Philipp Mehl, Richard Boire, Robert Lepenies, Ronny Röwert, Sebastian Moraga Scheuermann, Thorsten Thiel, Tony Ross-Hellauer, Vince I. Madai, Vincent Traag, Wojciech Hardy, Zining Zhu. We also extend our thanks to the 9 experts who preferred to remain anonymous. We furthermore would like to extend our appreciation to our colleagues from the Global Network of Internet and Society Research Centers (https://networkofcenters.net/) for their invaluable assistance in the recruitment of experts and their insightful contributions in shaping initial ideas.

# APPENDIX

*Table 1. Opinions*

|  | Strongly disagree | Disagree | Neither agree nor disagree | Agree | Strongly agree |
|---|---|---|---|---|---|
| ChatGPT or other large language models will be widely used in administrative work. | 0.0 % | 7.7 % | 5.8 % | 53.8 % | 32.7 % |
| ChatGPT and other large language models amplify the problem of disinformation. | 3.8 % | 1.9 % | 19.2 % | 42.3 % | 32.7 % |
| ChatGPT and other large language models will transform scientific work. | 0.0 % | 7.7 % | 21.2 % | 44.2 % | 26.9 % |
| ChatGPT and other large language models need to be more regulated. | 7.7 % | 15.4 % | 13.5 % | 25.0 % | 38.5 % |
| I will definitely use ChatGPT or other large language models in my scientific work. | 1.9 % | 13.5 % | 25.0 % | 32.7 % | 26.9 % |
| The advantages of ChatGPT and other large language models for the science system outweigh their disadvantages. | 5.8 % | 15.4 % | 28.8 % | 36.5 % | 13.5 % |
| ChatGPT and other large language models will make science even more dependent on commercial vendors. | 3.8 % | 19.2 % | 30.8 % | 36.5 % | 9.6 % |
| ChatGPT and other large language models should be able to be listed as co-authors in the texts they help produce. | 40.4 % | 17.3 % | 17.3 % | 15.4 % | 9.6 % |
| ChatGPT and other large language models will replace scientists in the long-run. | 59.6 % | 23.1 % | 9.6 % | 5.8 % | 1.9 % |

*Table 2. Applications*

| | r1 | r2 | r3 | sum | rank | rank_perc |
|---|---|---|---|---|---|---|
| Applications | | | | | | |
| text improvement (e.g. with regard to spelling, grammar, style and expression) | 17 | 5 | 9 | 31 | 87 | 23.9 |
| text summary (e.g. reduce texts to key points) | 8 | 19 | 15 | 42 | 85 | 23.4 |
| code writing (e.g. write programs) | 9 | 7 | 13 | 29 | 63 | 17.3 |
| idea generation (e.g. creating new approaches, linking concepts) | 10 | 7 | 2 | 19 | 56 | 15.4 |
| text translation(e.g. transfer texts from one language to another) | 6 | 7 | 3 | 16 | 41 | 11.3 |
| text analysis (e.g. identify specific features within a text such as arguments and claims) | 2 | 7 | 10 | 19 | 32 | 8.8 |

*Table 3. Limitations*

| | r1 | r2 | r3 | sum | rank | rank_perc |
|---|---|---|---|---|---|---|
| Limitations | | | | | | |
| incorrectness (e.g. quotes, literature references, biographies, etc. are invented) | 25 | 11 | 9 | 45 | 131 | 36.0 |
| lack of transparency (e.g. unclear where data comes from and what sources are used) | 11 | 19 | 14 | 44 | 96 | 26.4 |
| unspecificity (e.g. produce superficial texts that do not address topics in depth or detail) | 10 | 17 | 15 | 42 | 89 | 24.5 |
| outdatedness (e.g. draws on outdated data) | 4 | 4 | 6 | 14 | 30 | 8.2 |
| not-creativity (e.g. limits creativity in writing) | 2 | 1 | 8 | 11 | 18 | 4.9 |

Table 4. Opportunities

| | r1 | r2 | r3 | sum | rank | rank_perc |
|---|---|---|---|---|---|---|
| Opportunities | | | | | | |
| reduce administrative tasks (e.g. reports can be written semi-automatically) | 13 | 15 | 18 | 46 | 100 | 27.5 |
| promote efficiency (e.g. researchers can produce texts faster) | 10 | 19 | 10 | 39 | 88 | 24.2 |
| promote inclusiveness (e.g. by helping researchers overcome language limitations) | 13 | 7 | 6 | 26 | 72 | 19.8 |
| promote productivity (e.g. researchers can produce more analyses or more articles) | 8 | 7 | 12 | 27 | 58 | 15.9 |
| promote reflection (e.g. requires a critical analysis of correct and incorrect statements which are presented equally convincingly) | 8 | 4 | 6 | 18 | 46 | 12.6 |

Table 5. Risks

| | r1 | r2 | r3 | sum | rank | rank_perc |
|---|---|---|---|---|---|---|
| Risks | | | | | | |
| reinforce bias / dominant voices (e.g. large language models primarily reproduce majority positions) | 18 | 8 | 5 | 31 | 93 | 25.5 |
| lead to an increase in disinformation (e.g. erroneous scientific content is difficult or impossible to distinguish from accurate scientific content) | 8 | 8 | 19 | 35 | 67 | 18.4 |
| overburden academic quality assurance mechanisms (e.g. a large numbers of machine-generated articles will be submitted to journals) | 8 | 8 | 6 | 22 | 54 | 14.8 |
| encourage academic misconduct (e.g. the bar for misuse will be lowered) | 5 | 10 | 8 | 23 | 48 | 13.2 |
| lead to a decrease in originality (e.g. produced texts rely on existing information) | 6 | 6 | 5 | 17 | 41 | 11.3 |
| increase dependence on commercial providers (e.g. quality models will more likely be created by commercial actors than by non-commercial actors) | 5 | 3 | 7 | 15 | 33 | 9.1 |
| reinforce inequalities (e.g. between individuals who can afford to pay for such tools and those who cannot) | 2 | 9 | 2 | 13 | 28 | 7.7 |

Table 6. Competencies

| | r1 | r2 | r3 | sum | rank | rank_perc |
|---|---|---|---|---|---|---|
| Competencies | | | | | | |
| reflective mindset (e.g. critically evaluate the influence of large language models on research) | 18 | 17 | 9 | 44 | 115 | 31.6 |
| ability to contextualize results (e.g. interpret results and their feasibility in practice) | 20 | 11 | 12 | 43 | 114 | 31.3 |
| ethical understanding (e.g. being aware of the responsible use of AI) | 6 | 13 | 21 | 40 | 71 | 19.5 |
| technical know-how (e.g. understanding how large language models work) | 8 | 11 | 10 | 29 | 64 | 17.6 |

Table 7. Codebook Phase 1

| main codes | definition | example quote |
|---|---|---|
| applications: use in scientific work | | |
| text improvement | statements about the improvement, rephrasing / polishing, translation and optimization of text. | "assisting with basic writing and formatting. In short anything where the user knows the correct answer, but lacks the time to write an entire paragraph about it for the 100th time." |
| text summary | Statements about summarizing or simplifying a text from an input to a LLM. | "LLMs are a great tool for helping writers to overcome "the white page syndrome". Just ask the LLM to write an intro paragraph about a subject, and it should be enough to give the writer inspiration to proceed with the narrative. Also, scientific writing is often done in a hurry and scientific papers often lack a good presentation or writing form. LLMs can be used to help writers to refine critical portions of text, such as abstracts and summaries." |

| main codes | definition | example quote |
|---|---|---|
| text analysis | statements about the usage of a LLM for the analysis, guidance and suggestions of data. | "Tatsächlich aber auch, um Lücken in Protokollen etc. zu finden. Beispiel: "Take this protocol as an input. What other research questions would you suggest?" |
| code writing | statements on the assistance with coding including offering auto-completion, guidance, error troubleshooting, explanations of coding concepts and terminology. | "1- Help developing code.<br>2- Help debugging code.<br>3- Help learning new coding packages and languages." |
| idea generation | statements on generating new ideas by offering or structuring insights, suggestions, and contextual information. | "Support in coding, support in writing and editing, documentation (e.g. of codes), coming up with titles, writing abstracts/summaries. Brainstorming ideas (on hypotheses, research directions, literature gaps, etc.)." |
| text translation | statements about the application of a LLM as a translating device as well as their integration in existing applications for translation. | ""ChatGPT might be able to provide useful draft texts for non-essential writing, such as emails or invitation letters.<br><br>An additional option might be to let ChatGPT edit texts to improve readability. This could be helpful in various ways. For instance, it can easily put parts of texts in different tenses or voices or even different styles." |
| **limitations: use in scientific work** | | |
| lack of transparency | statements on the perpetuation or amplification of biases present in the training data and lack of transparency of LLMs as well as how they arrive at their outputs /. Relates to the generation of the output | "The biggest danger is in using them to summarise information - difficulties in their ability to attribute answers, risks of harm of incorrect information disseminating. Biases in training data probably strictly limit their perspectives. The cost in terms of dollars, and to the environment in terms of energy consumption also likely prohibit broader use (when not subsidised by tech companies eager for growth)" |
| incorrectness | statements about the outputs that are false or misleading based on data or language patterns that are outside of an LLM's training or understanding / Relates to the output. | "ChatGPT seems to produce plausible results, but at the moment in my experience it can mess up quite spectacularly while looking plausible" |
| non-creativity | statements about the reliance on existing patterns and language structures in an LLM's training data and the difficulties to generate entirely new or creative content without significant human input or manipulation | "Answers often depend significantly on asking the question in the "right" way; Since it is based on existing knowledge, it can only ever create something that has already been thought of" |

| main codes | definition | example quote |
|---|---|---|
| outdatedness | statements about the inaccuracy of data over time as language patterns and cultural contexts evolve as well as (un)relevance and usefulness of outdated data. | "Could help in polishing an idea? However, this is limited as well as ChatGPT is not continuously updated as in Google search. Data for it is only until 2021!" |
| unspecificity | statements about LLMs producing vague or ambiguous outputs that lack specificity or clarity reflect the concern over instances where the generated content fails to provide precise or unambiguous information as well as statements about generic or generalized responses without addressing the specific nuances of a given query or context. | "Die Ergebnisse sind eher oberflächlich, sie bilden die weitläufige Meinung zu einem Thema ab aber enthalten keine differenzierte Abwägung von Argumenten. Zudem die bekannten Kritiken, wie Validität der Aussagen und Überprüfbarkeit der Quellen." |
| opportunities for the science system | | |
| efficiency | statements about the improvement of speed and accuracy of scientific research and analysis by automating language-based tasks, such as data extraction and analysis, literature reviews, and scientific writing. | "Die Hoffnung wäre Schreibproduktivität in Bereich auf Forschung (oder Effizienverbesserung in Bezug auf den großen Bereich von wissenschaftsbegleitenden Tätigkeiten), wie immer mit dem Risiko, dass dann einfach Erwartungen und Prozesse weiter hochgeschraubt werden und man vor allem auch noch mehr zu lesen vorgehalten bekommt..." |
| reflection | statements about the reflection of research methodologies and assumptions due to potential biases or gaps in existing data and analysis, identification of new research areas and directions based on the analysis of existing research and data as well as rethinking and renewing science due to the opportunities as well as the improvement of productivity and efficiency. | "Das Wissenschaftssystem an sich könnte ChatGPT nutzen, um Biases, Diskriminierungen, Tendenzen usw. im eigenen System festzustellen." |
| reduce administrative tasks | statements about reducing administrative tasks in a codebook highlight the potential benefits of leveraging Large Language Models (LLMs) to streamline and simplify various administrative processes. This includes automating data entry, document management, scheduling, and communication tasks, among others. | "Einen echten Wert sehe ich auch noch in den Möglichkeiten, administrative Breiche des wissenschaftlichen Betriebs stärker zu automatisieren, insbesondere das Berichtswesen und ggf. auch Vorarbeiten zu Wissenschaftskommunikation" |

| main codes | definition | example quote |
|---|---|---|
| inclusiveness | statements about advancing equity such as facilitation of the inclusion of underrepresented groups in research and scientific communication through the use of more accessible and inclusive language, and eliminating language barriers and biases that can impede the participation and contribution of individuals from diverse linguistic and cultural backgrounds in scientific research and collaboration, leading to a more equitable and diverse scientific community. | *"I agree with the examples provided. In addition: scaling research; generating ideas beyond the thinking of research teams; possibly levelling of the playing field across institutions and/or countries"* |
| productivity | statements about the productivity of scientists due to automation of language-based tasks including literature reviews, data extraction, and scientific writing, enable time for critical thinking, analysis, and experimentation. | *"Increase in productivity. But we need to use them correctly -- their generated contents shouldn't be considered as truths. We shouldn't rely on these contents."* |
| **risks for the science system** | | |
| overburden academic quality assurance mechanisms | statements about the production of an overwhelming amount of scientific papers, data sets, and findings. | *""Steigerung der produzierten Texte nicht wünschenswert, da bereits ohne ChatGPT die Menge und Breite der publizierten Texte in vielen Bereichen unüberschaubar ist."* |
| inequalities | statements about the disproportionately benefit of scientists with better access to technology and resources, further widening the digital divide and exacerbating existing inequalities in the scientific community. | *"Fairness: who has access (small universities in the global south having a different acces compare to Google researchers?). Also: will it be available in many languages or will it just enforce English?"* |
| dependence | statements about the reliance on technology and automated processes that could be vulnerable to technical issues and malfunctions, potentially leading to delays, errors, and inaccuracies in scientific research and | *"Abhängigkeit von KI-Anbietern, Intransparenz von Forschungsprozessen durch Blackbox KI"* |

| main codes | definition | example quote |
|---|---|---|
| | analysis. Also statements on the dependence on proprietary software and platforms, limiting access to and control over scientific data and findings. | |
| misconduct | statements about the intentional or unintentional misconduct of LLM's, such as the usage of manipulated data or the dissemination of false or misinterpreted information through individual scientific misconduct. | "promote unvetted information" |
| decrease in originality | statements on the production or reinforcement of existing biases and paradigms, including the lack of diverse perspectives and novel approaches, discouragement of critical thinking and independent analysis as well as conformity and acceptance of existing norms (rather than challenging them). | "Mangelnde Kreativität. ChatGPT et al werden keinen relevanten kreativen Erkenntnisfortschritt erreichen. Dennoch können sie in einer weitgehend Paradigmen-diferenzierten Wissenschaft durchaus niveaugleiche Texte produzieren. Was nicht notwendig an der guten Qualität der KI-gestützten Texte liegt, sondern in einem qualitätsverfall wissenschaftlicher Texte." |
| reinforce bias / dominant voices | statements on a LLM's potential to unintentionally or intentionally perpetuate biases or favor dominant voices as well as mainstream voices within the generated outputs. | "Main risk is the flooding of media with shallow and non-checked information, which can contribute to higher misinformation and lower trust in science." |
| disinformation | statements about the potential risk of unintentionally or intentionally spreading false or misleading information through the use of LLMs that may be trained on biased or inaccurate data, leading to the spread of disinformation and misinformation, ultimately eroding the credibility of scientific institutions and research, and decreasing public trust in the scientific community regarding the practice by scientists. | "Darüber hinaus ist außerhalb der Wissenschaft die gefahr groß, dass wissenschaftliche Informationen ohne gründliche Recherche für die Öffentlichkeit aufbereitet werden und es zu Desinformation kommt." |
| competencies in usage | | |

| main codes | definition | example quote |
|---|---|---|
| technical know-how | statements about the technical understanding of LLMs, such as its strengths, limitations, potential biases, and implications. | *"Basic technical knowledge of the inner works"* |
| ethical understanding | statements about the ethical justifiability of using LLMs such as the perpetuation of biases, lack of transparency in decision-making processes, and the potential risk of unintended consequences, including amplifying existing inequalities or creating new ones. | *"A legal problem with ChatGPT is the way that authorship and copyright are attributed. This holds for both, the texts that ChatGPT is trained on and the texts that are produced with the help of ChatGPT. A moral problem has to do with the ethics of research. Even before AI, there were many instances where researchers could choose to omit certain actions that are morally obligatory in science. Now that the production of text can be (partially) exported to an AI, it is easy to give in to the temptation of not caring too much whether the wording provided by some AI is accurate or not. Of course, the problem intensifies in a context where researchers are pressured to publish."* |
| reflective mindset | statements of the need to understand the implications of the use of LLM for scientific work and the broader science system, including issues related to bias, transparency, accuracy, originality, productivity, and ethics, as well as consideration of how to mitigate potential risks and maximize the benefits of LLMs: What is the tool doing with us? | *"Motivation for reflecting on what researchers actually do"* |
| ability to contextualize results | statements about the scientist's ability to apply and embed the results of LLMs into the specific context of their research or scientific work, ensuring that the generated insights are relevant and meaningful for their particular field or domain: What are we doing with the tool and output? | *"Eingabekompetenz, d.h. die Fähigkeit "prompts" so zu formulieren, dass die Antwort möglichst genau/differenziert ist. Bewertungskompetenz, d.h. Kontextualisierung der Ergebnisse."* |
| legal implications for scientists when using LLMs | | |

| main codes | definition | example quote |
|---|---|---|
| copyright | statements about outputs that infringe copyright laws or intellectual property rights, usage of copyrighted or proprietary data without proper authorization or reference. | "Copyright might become an issue if the model does not alter its source material in a sufficient manner; There is an incentive for basically everyone to try and find out how far they can take the technology, increasing the likelihood of fraudulent behaviour; Referencing is usually not happening within ChatGPT, so it does not cite the sources it makes use of, which is at least problematic and at worst could result in plagiarism;" |
| data protection | statements about complying to privacy laws, such as ensuring the confidentiality and protection of personal data used to feed models, and compliance with regulations when it comes to sensitive personal data. | "I see huge problem with copyright and data protection because OpenAI has not made it transparent where the data is coming from, how much data was used and if the rightful owners of the data agreed to its use in training ChatGPT. This is especially a problem for already marginalzed groups. Also there is the risk that bias that already exists in the data used to train ChatGPT is reproduced in the system and goes unchecked." |
| liability | statements about a Scientist's potential liability for any unintended consequences or errors that may arise from the use of LLM's. | "[Verbreitung] (Rassistische[r] und sexistische[r] Ergebnisse)" |
| **ethical implication for scientists when using LLMs** | | |
| need for accountability in relation to the outcomes produced by LLMs | statements about the accountability for the results obtained from using LLMs including to ensure critical analysis or evaluation of outputs regarding ethical implications. | "- Accountability is a big issue if ChatGPT is accepted as a coauthor. Are machine can hardly be held responsible. - Chatbots may spread stereotypes and biases." |
| originality with regards to human creativity | statements about authorship, credit, plagiarism regarding the output generated by LLMs. | "- plagiarism - definition of authorship - legal: author(ity)" |
| sustainability | Statement about an LLM's environmental impact from their development to usage, such as high computational resources that result in carbon footprint and energy consumption. | "The main problems that I see are the provision of incorrect answers by LLMs, and the risks associated with relying on commercial products and intransparent algorithms. Other issues include the question of costs and the consequences of unnecessary use of ChatGPT or other similar tools with regard to energy consumption." |
| potential exclusion of researchers | statements about the implications of potential for unequal access and high costs limiting the ability of some scientists to use these tools and contribute to scientific knowledge. | Fairness: who has access (small universities in the global south having a different acces compare to Google researchers?). Also: will it be available in many languages or will it just enforce English? Autorship, copyright: these are legal questions. My opinion is that these two point require a human, with moral and legal agency (i.e. legal responsibility). |

| main codes | definition | example quote |
|---|---|---|
|  |  | *Data protection is interesting. Many people will not think twice and upload all their data to an LLM, asking for "write me a paper", despide any confidentiality or ethical constraints linked to the data.* |
| issue of autonomy | statements about dependencies of scientists. | *"Ethisch - in einem umfassenden Sinne - besteht die große Gefahr der Verkümmerung menschlicher Fähigkeiten. Dies ist aber nicht neu oder disruptiv durch ChatGTP, sondern durch alle ubiquitären Softwaresyteme der Fall. (Kurze schnelle Recherche, Sprachoptimierung (englisch), kurze u, Pos. 6)"* |
| **future perspectives for the science system** |  |  |
| transformation | statements about the transformation of the science system towards such as interdisciplinary collaboration, open access and transparency, and responsible use of AI for scientific research. | *"It could make certain scientific tasks, like literature research, a thing of the past, if the model could have meaningful access to the current state of research;*<br>*Certain types of publications in specific areas could be fully automatized, like literature analysis;*<br>*It could open up many additional languages to researchers, as translation could work seamlessly and with high quality outputs;*<br>*New ways of doing science might spring up, it might be more important to be able to configure the language models themselves in the future than analyze their results yourself;"* |
| deformation | statements about the deformation of the science system such as unequal distribution of knowledge and resources, exacerbating existing inequalities and creating new ones, and undermining the credibility and trustworthiness of scientific research. | *"Die Gefahr besteht in einer qualitativen Nivellierung wissenschaftlichen Outputs auf mittelmäßigen Niveau. Da dies kollektive Phänomene sind, wird das individuell aber nicht bemerkbar sein. D.h. es wird weder in der Wissenschaftscommunity noch der Öffentlichkeit wirklich wahrgenommen werden. Wenn das System Wissenschaft sich dieser Problematik nicht autoregulativ stellt, dann wird sie auf kurz oder lang in vielen Bereichen irrelevant werden und nur noch autopoeitisch operieren. Da die Immunisierungstendenzen ohnehin schon quer in und durch alle Fachbereiche bemerkbar sind, ist meine Prognose pessimistisch. Wissenschaft wird weiterhin existieren, aber die gesellschaftliche Relevanz wird sinken, sofern sie sich nicht selbst intern herausfordert. Dieser Qualitätsverlust ist durch heutige Kuhnsche Paradigmen-Wissenschaft ohnehin schon bemerkbar, allerdings ohne in eine fundamentale Krise zu geraten. Sollte sich dieser Trend verstetigen, dann werden sich Wissenschaft und andere gesellschaftliche Systeme weiter disjunktiv trennen ohne in eine funktionale Differenzierung zu geraten. Ergo: Die Gesellschaft nutzt ChatGTP und andere zur Information und Wissensgenerierung und die Universitäten verleihen Hochschulabschlüsse als Status- und Qualitifkationssignal sowie die Professoren und andere wiss. Tätige bespielen ihre eigene Bubble. Diese Einschätzung ist pessimistisch, aber durchaus auch realistisch."* |
| no significant change | statements about the science system remaining unchanged | *"ChatGPT is like another programming language. I don't think the scientific practices will be fundamentally changed by another programming language."* |

*Table 8. Codebook Phase2*

| scenario | main Codes | definition | example quote |
|---|---|---|---|
| deformation | **negative impact on scientific infrastructure/eco system** | | |
| | replacement of support posts / administrative jobs | statements about the potential risk of LLMs replacing or reducing the need for certain support posts or administrative jobs that involve tasks such as data entry, transcription, or document analysis | *"Daraus folgend werden ggf. weitere Berufsgruppen (Sekretariat, Verwaltungen, Werbetexter, Callcenter/Kundenservice, etc.) weiter automatisiert und entsprechende Jobs verschwinden. Diese Freisetzung von Menschen wird (hoffentlich) zu einer Einführung des Grundeinkommens führen :)"* |
| | reinforce inequalities | statements about the potential risk of LLMs perpetuating or amplifying limiting access to LLM resources or knowledge, thereby exacerbating the digital divide and marginalizing disadvantaged or underrepresented groups | *"On the negative side, society will generally become more competitive and probably add a massive gap between those who know how to use Ai and those who don't."* |
| | devaluation of science | statements about the potential of a reduction in the perceived importance, credibility, or authority of scientific knowledge and expertise | *"Vielleicht wird die Anerkennung für wissenschaftliches Arbeiten und Textarbeit geringer."* |
| | generate dependencies | statements about the potential risk of LLMs creating a reliance on proprietary software or commercial (third-party) providers for access to LLM resources, tools, and expertise, which may limit academic freedom, restrict innovation, and raise ethical concerns related to ownership, control, and transparency of research outputs. | *"Abhängigkeiten von kommerziellen Anbietern"* |
| | **negative impact on scientific quality** | | |
| | reinforce predatory publishing | statements about the potential risk of LLMs being used to generate low-quality or fraudulent content, which may be published in predatory journals for financial gain, thereby compromising the integrity and credibility of scientific research and contributing to the spread of misinformation. | *"The biggest potential problem I see is that if there's a greater deluge of predatory publishing practices being driven by LLM"* |
| | decrease of quality | statements about the potential negative impact on the quality of research or outputs resulting from overreliance on LLMs without proper scrutiny, which can lead to errors, biases, or other limitations in the data or analysis. | "There is a risk on lowering quality through ran increased redundant replication of existing work and ideas though." |
| | homogenization of science | statements about the potential risk of reducing diversity and creativity in scientific research or outputs, as reliance on LLMs may lead to the replication of existing knowledge and biases, mainstream voices, limiting the exploration of new and diverse research questions or approaches. | "They also draw on existing corpuses of knowledge which are biased towards certain perspectives. Bias is inevitably something scientists deal with on a regular basis, but without knowing the algorithms underlying AI models, it is much more difficult to assess bias." |
| | loss of intellectual ability to think | statements about the potential risk of decreasing the intellectual independence and critical thinking skills of researchers or users who over-rely on LLMs, leading to a loss of creativity and originality in research or outputs | "There could be the risk that researchers lose the ability to think about their results and claims made thoroughly and to construct new ways of argumentation and classifcation." |
| | **negative impact on scientific integrity** | | |
| | undermine academic training | statements about the potential risk of replacing traditional academic training and critical thinking skills with an over-reliance on LLMs for tasks such as data analysis or interpretation, which can lead to a lack of understanding of underlying principles and limitations and undermine the development of essential academic and professional skills | "Problematisch ist dabei, dass (die meisten) jungen Wissenschaftler:innen als Teil ihrer Ausbildung diesen repititiven [sic] Prozess brauchen, um Basiswissen parat zu haben und eben auch die Leistung der KI-generierten Texte einschätzen zu können ("repetitio est mater studiorum" - ist was dran)." |
| | loss in science credibility | statements about the potential risk of reducing the credibility of scientific research or outputs, as over-reliance on LLMs without proper scrutiny or critical evaluation can lead to errors, biases, or inaccuracies, thereby diminishing the trust and confidence in scientific findings | "A main risk that I see, however, is the potential of an increased generation and proliferation of misinformation. If AI tools are, e.g., abused to "mass-produce" papers of low quality or containing misinformation this could, ultimately, lead to a decreased trust in science." |
| | spreading disinformation | statements about the potential risk of unintentionally or intentionally spreading false or misleading information through the use of LLMs that may be trained on biased or inaccurate data, leading to the spread of disinformation and misinformation | "On the other hand, I think AI can encourage scientists to be lazy, both in producing new ideas and in being critical of the ideas of others. It increases the risk of disinformation, since AI models can invent sources, as we have seen." |
| | **positive impact on scientific creativity and discovery:** | | |

**transformation**

| | | | |
|---|---|---|---|
| | augment science | statements about the potential for LLMs to enhance and advance scientific research and outputs by enabling the processing and analysis of vast amounts of complex data, supporting the exploration of new research questions, and promoting greater efficiency and accuracy in scientific workflows. | *"The most exciting change will be the ability to link ideas and research on a much greater scale than is currently possible. Ideally it will also speed up the time between experiment and publication and also reduce administration, allowing researchers to focus on science."* |
| | **positive impact on science education and communication:** | | |
| | Enhance science | statements about the potential for LLMs to improve and strengthen scientific research and outputs by providing new insights and perspectives, facilitating interdisciplinary collaborations, supporting the development of innovative research methods, and advancing the overall understanding of complex phenomena in science and beyond | *"The positive potential of AI in academia is that it can be used to increase the efficiency of research and, ideally, also contribute to the quality of its products (publications, data, code/software). In an ideal scenario, this could also contribute to increasing the appreciation of as well as the trust in science."* |
| | innovate education | statements about the potential for LLMs to transform and improve education by providing new opportunities for personalized learning, enabling the development of intelligent tutoring systems and educational chatbots, enhancing the accessibility of educational resources, and promoting greater engagement and motivation among students | *"I believe consequences will be majorly positive as it will make the process of learning much easier than before."* |
| | **positive impact on scientific productivity and efficiency** | | |
| | streamline tasks | statements about the potential benefit of LLMs in automating and optimizing certain tasks related to language processing, such as language translation, text summarization, or sentiment analysis, which may save time and resources, improve accuracy and consistency, and enhance the efficiency and effectiveness of various domains, such as healthcare, education, or customer service. | *"In my opinion generative AI will have an effect on administrative and formal tasks by writing applications, assesments proposals, abstracts or status reports, basically everything which is highly standardized and therefore easy to replicate"* |
| | assistance | statements about the potential benefit of LLMs in providing intelligent and personalized assistance to users in various domains, such as language learning, writing, or content creation, by offering suggestions, corrections, or feedback, which may improve the quality and accessibility of education and communication, as well as enhance the user experience and satisfaction but also in other areas such as public health etc. | *"Ich denke, dass man die Sprachmodelle weniger zur Generierung von Inhalten nutzen wird, sondern mehr als Suchmaschine, mit der man in Interaktion treten kann."* |